\documentclass[iop]{emulateapj} 
\usepackage{natbib,amsmath,booktabs,apjfonts,threeparttable,multirow}

\usepackage{graphicx}
\usepackage{gensymb}
\usepackage{epstopdf}
\usepackage{color}

\newcommand{\msh}{MSH 17--3{\it 9}}

\newcommand{\ctbt}{G337.0--0.1}

\newcommand{\fermi}{{\it Fermi}-LAT}

\newcommand{\G}{G5.71--0.08}
\newcommand{\Gg}{G5.7--0.1}
\newcommand{\g}{G5.7--0.0}
\newcommand{\FGL}{2FGL J1758.8--2402c}

\shortauthors{Joubert, Castro, Slane \& Gelfand}
\slugcomment{Accepted for Publication in The Astrophysical Journal}

\begin{document} 
\title{{\it FERMI}--LAT Observations of Supernova Remnant {\Gg}, Believed to be \\Interacting with Molecular Clouds}
\author{Timothy Joubert\altaffilmark{1,4}, Daniel Castro \altaffilmark{2}, Patrick Slane\altaffilmark{1}, Joseph Gelfand\altaffilmark{3}}

\altaffiltext{1}{Harvard-Smithsonian Center for Astrophysics, 60 Garden Street, Cambridge, MA 02138, USA}
\altaffiltext{2}{MIT-Kavli Center for Astrophysics and Space Research, 77 Massachusetts Avenue, Cambridge, MA, 02139, USA}
\altaffiltext{3}{NYU Abu Dhabi, PO Box 129188, Abu Dhabi, UAE}
\altaffiltext{4}{Student Squadron, 47th Flying Training Wing, Laughlin AFB, TX, 78843, USA}
\begin{abstract}

This work reports on the detection of $\gamma$-ray emission coincident with the supernova remnant (SNR) {\Gg} using data collected by the Large Area Telescope aboard the {\it Fermi Gamma-ray Space Telescope}. The SNR is believed to be interacting with molecular clouds, based on 1720 MHz hydroxyl (OH) maser emission observations in its direction. This interaction is expected to provide targets for the production of $\gamma$-ray emission from $\pi^0$-decay.  A $\gamma$-ray source was observed in the direction of SNR {\Gg}, positioned nearby the bright $\gamma$-ray source SNR W28. We model the emission from radio to $\gamma$-ray energies using a one-zone model. Following consideration of both $\pi^0$-decay and leptonically dominated emission scenarios for the MeV-TeV source, we conclude that a considerable component of the $\gamma$-ray emission must originate from the $\pi^0$-decay channel.  Finally, constraints were placed on the reported ambiguity of the SNR distance through X-ray column density measurements made using XMM-Newton observations. We conclude {\Gg} is a significant  $\gamma$-ray source positioned at a distance of $\sim 3$ kpc with luminosity in the 0.1--100 GeV range of $L_{\gamma} \approx 7.4 \times 10^{34}$ erg/s.
\end{abstract}


\section{INTRODUCTION}
\label{sec:Introduction}

Observational evidence supporting the diffusive shock acceleration (DSA) of cosmic rays (CR) has steadily increased over the last few decades. Current evidence includes observations of nonthermal X-ray emission from shell-type SNRs, believed to be the result of synchrotron radiation from shock accelerated electrons \citep{Lazendic_2004, Vink_2012}. Also, very high energy (VHE) $\gamma$-ray emission has been observed in the direction of a significant number of SNRs, including SN 1006, RX J0852.0-4622, and IC 443 using ground based Cherenkov telescopes \citep{Aharonian_2007a, Acciari_2009, Berezhko_2009}. Both the X-ray and VHE $\gamma$-ray cases support a nonthermal shock acceleration mechanism, one that would result in acceleration of both hadrons and electrons, which can be termed hadronic and leptonic acceleration respectively. There are several emission mechanisms that result from the acceleration of electrons including nonthermal bremsstrahlung emission and inverse Compton scattering (IC) of photons by accelerated electrons. Hadronic acceleration allows for a separate $\gamma$-ray emission mechanism caused by interactions between the accelerated hadrons and ambient material \citep{Ellison_2010}. 

The hadronic mechanism should theoretically produce $\gamma$-ray emission through the decay of neutral pions, $\pi^0 \rightarrow \gamma \gamma$, produced by proton-proton, generally nuclear-nuclear, interactions of the shock accelerated ejecta and ambient material \citep{Ackermann_2013}. This material often exists in the form of molecular clouds (MC) adjacent to the SNR \citep{Drury_1994, Claussen_1997}. Acceleration of cosmic-ray protons into these clouds, or re-acceleration of ambient cosmic rays, is believed to enhance the $\pi^0$-decay $\gamma$-ray emission signature due to more frequent pion producing interactions because of the higher proton density relative to the ISM \citep{Aharonian_1994, Drury_1994, Uchiyama_2010, Tang_2014, Lee_2015}. Therefore, observations of SNR and molecular cloud interaction, often termed SNR-MC systems, provide a particularly interesting astrophysical laboratory for studying the signatures of a hadronic $\gamma$-ray production mechanism \citep{Yusef_2002}. 

The theorized peak of the resulting photon spectrum is well positioned within the detectable energy range of the $\emph{Fermi Gamma-ray Space Telescope}$ Large Area Telescope (\fermi), and is characterized by a sharp rise between $\sim70 - 200$ MeV \citep{Kamae_2006, Reynolds_2008}. {\fermi} has made significant improvements in sensitivity and resolution over previous instruments covering this energy range, most notably the Energetic Gamma-Ray Experiment Telescope (EGRET). This makes {\fermi} study of SNR-MC systems very promising. Further, recent \fermi~observations of SNR-MC systems have corroborated theory, establishing these interactions as significant $\gamma$-ray sources, including correlation studies performed by \citet{Hewitt_2009}. In fact, SNR-MC systems are among the most luminous SNRs in $\gamma$-rays \citep{Abdo_706_2009, Thompson_2011}. 

While there are contrasting theories regarding the specific acceleration mechanism, evidence supporting a hadronic origin of observed $\gamma$-ray emission has recently strengthened. Observation of the SNR-MC systems IC443 and W44 by \citet{Ackermann_2013} clearly defined the $\pi^0$-decay signature down to 60 MeV, effectively demonstrating the hadronically dominated origin of the observed $\gamma$-ray emission in the direction of each SNR. Additional support for the connection of SNR-MC systems to hadronic acceleration comes from observations of the SNRs W41, \msh\ and \ctbt\ by \citet{Castro_2013}. This work compares three unique SNR-MC systems and the closeness of fit of two sophisticated $\gamma$-ray emission models to the broadband emission spectrum of each. The study reports that a model assuming a predominantly hadronic emission mechanism was in close agreement for the spectra of each SNR. Similar support is provided by the {\fermi} study of Kes 17 by \citet{Gelfand_2013} wherein a leptonically dominated emission scenario for the observed $\gamma$-ray emission was ruled out due to the high required electron-proton ratio as compared to local observations of the ratio. The {\fermi} study of SNR Kes 79 by \citet{Auchettl_2014} also supported a hadronic origin to the $\gamma$-ray emission based on both a high required electron-proton ratio compared to local observations, and also that the total electron energy required for a leptonically dominated emission origin would exceed the canonical SN explosion energy.

Evidence of SNR-MC interactions have been observed through focus on various emission lines including the detection of OH(1720 MHz) maser emission, which was first observed by \citet{Goss_1968} toward the SNRs W28 and W44. This emission has become a defining indicator of SNR-MC interaction. While other emission lines may indicate the existence of an MC within the field-of-view (FoV) of an SNR observation, the precise position of the MC is not easily determined. The OH(1720 MHz) maser emission line however is only emitted by dense gas ($\sim 10^5$ cm$^{-3}$) following the passage of non-dissociative shocks \citep{Lockett_1999}. This makes OH maser observations adjacent or interior to SNR radio and X-ray contours indicative of shock induced acceleration of cosmic ray hadrons, and potential sites of gamma-ray emission from these hadrons colliding with ambient material \citep{Claussen_1997, Koralesky_1998}.

The SNR W28 has been observed extensively at energies ranging from the radio to the very high energy (VHE) $\gamma$-ray regime. Within the field-of-view of W28 is the subject of this paper, SNR {\Gg} for which very little direct study has been conducted at any wavelength. {\Gg} was discovered by \citet{Brogan_2006} using observations of the the region taken with the VLA. Due to the SNR's relatively low radio brightness, {\Gg} was not definitively considered an SNR with a radio shell, and was not initially included in Green's catalog \citep{Green_2009}. The 2006 detection briefly established the existence of an SNR candidate named {\G} which was then changed to {\g} following additional studies of the region by \citet{Hewitt_2009}, using VLA and GBT observations. Data from the Multi-Array Galactic Plane Imaging Survey (MAGPIS) \citep{Helfand_2006} was also used to study the spatial characteristics of it's partial radio shell. This study confirmed the SNR categorization of {\Gg}.  

The SNR is believed to be interacting with neighboring molecular clouds, as evidenced by OH(1720 MHz) maser emission observed coincident with the radio center of the {\Gg} \citep{Hewitt_2009_OH} and also several studies of cloud interaction in the region of SNR W28 \citep{Claussen_1997, Arikawa_1999, Velazquez_2002}. The low radio brightness of {\Gg} has made previous attempts at setting constraints on the distance to the SNR difficult. Velocity measurements taken from OH(1720 MHz) maser observation set a kinematic velocity of 12.8 km~s$^{-1}$. Due to velocity ambiguity which could not be resolved, \citet{Hewitt_2009_OH} successfully set kinematic distance constraints at either 3.1 or 13.7 kpc. The maser observation establishes {\Gg} as a strong candidate site for shock induced cosmic-ray acceleration. In addition to maser detection, \citet{Aharonian_2008_W28} detected the VHE $\gamma$-ray source HESS J1800--240C inside the radio extent of the SNR while studying the W28 region, and also claimed a hadronic origin.

This paper will study the $\gamma$-ray emission in the direction of {\Gg} and seek to establish constraints on the nature of the production mechanisms by modeling the broadband spectrum. We will also use comparisons of different best-fit model parameters and XMM-Newton X-ray analysis of the region to set constraints on the distance to the SNR. A description of how the {\fermi} data were analyzed and the corresponding observation will be given in Section~\ref{sec:Observations} followed by a description of these results and their modeling in Section~\ref{sec:Discussion}. Finally a summary of the results is given in Section~\ref{sec:Summary}.

\begin{figure*}[!htb]
\minipage{0.5\textwidth}
\includegraphics[width=1.0\linewidth]{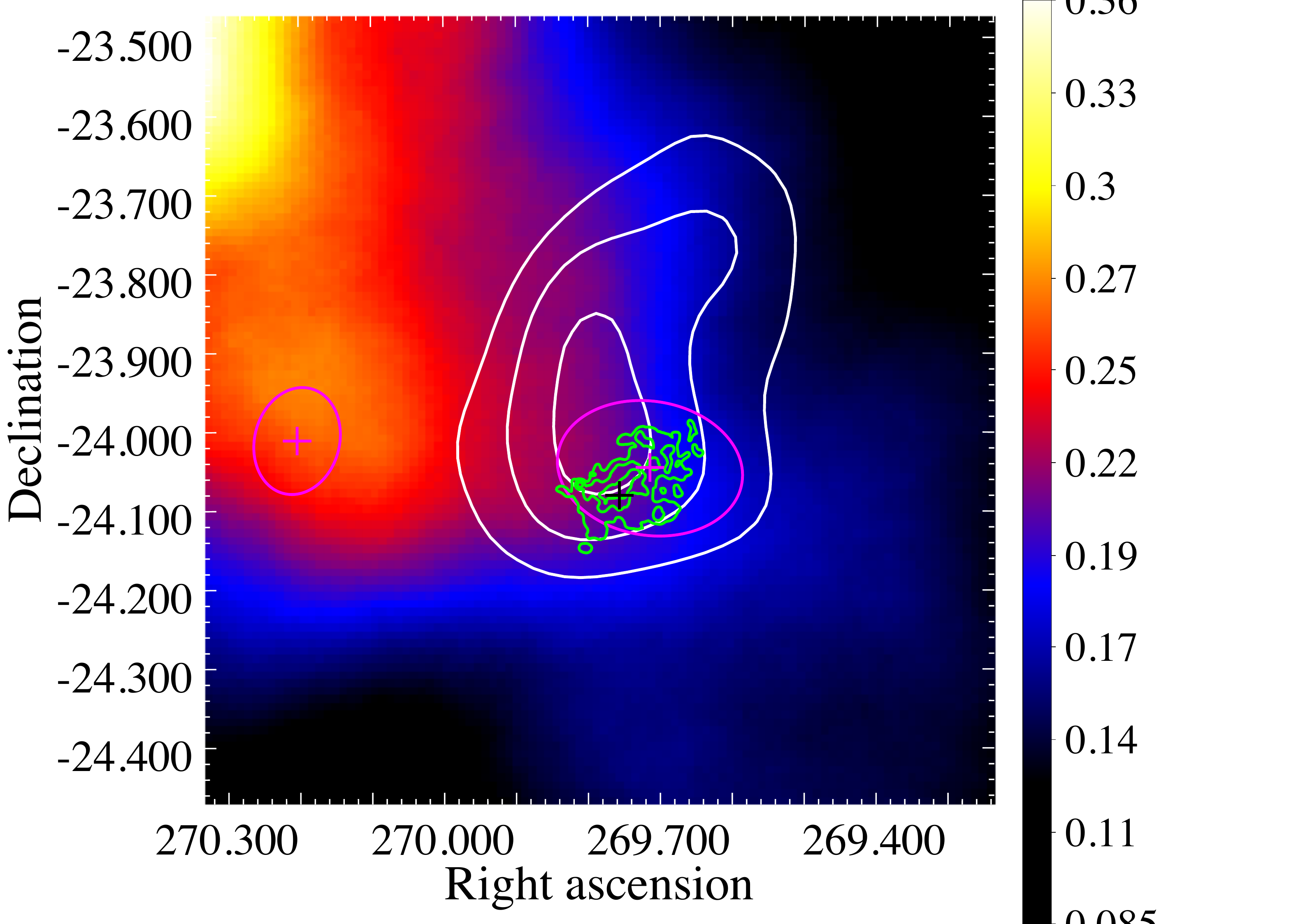}
\endminipage\hfill
\minipage{0.5\textwidth}
\includegraphics[width=1.0\linewidth]{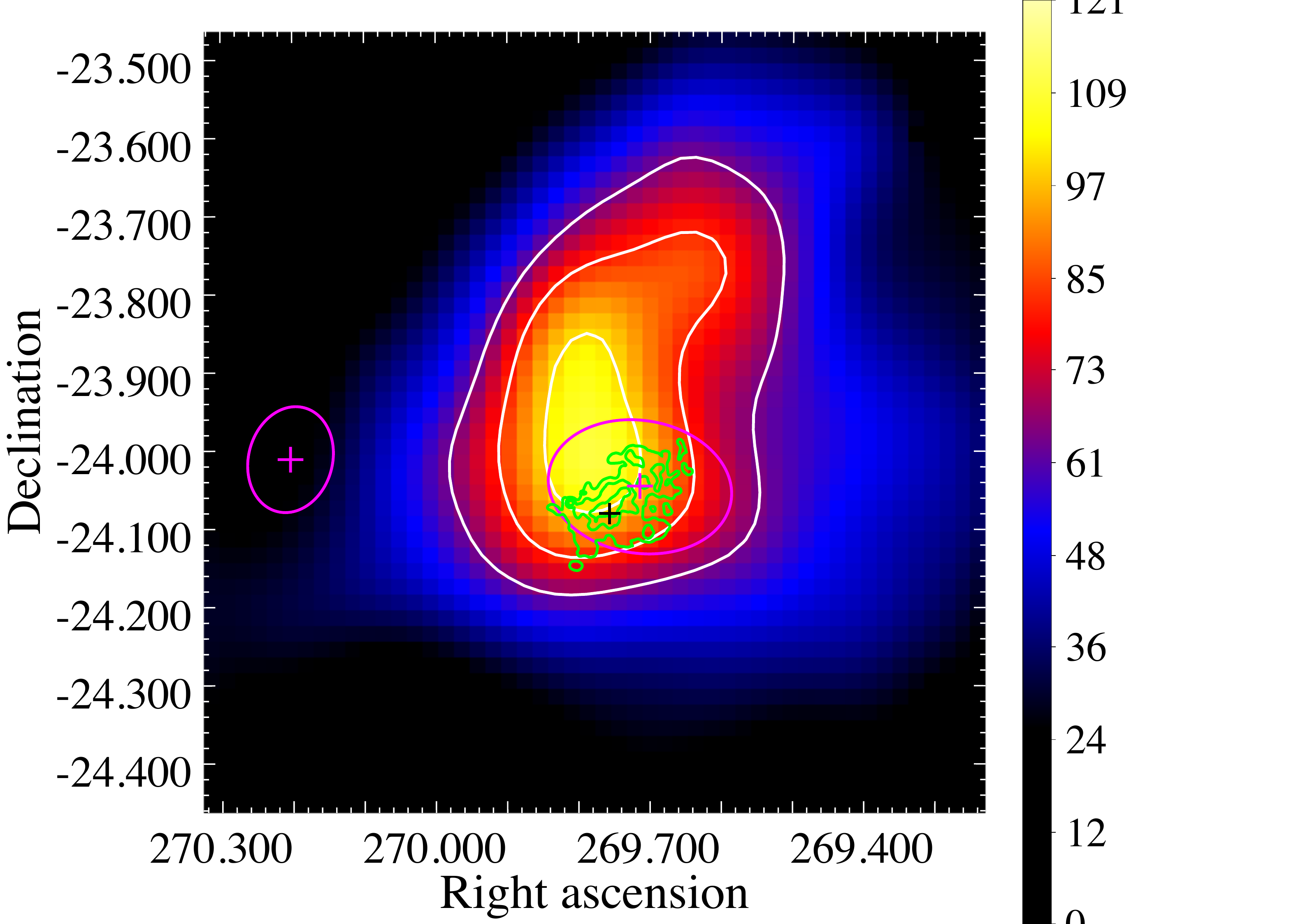}
\endminipage
\caption{The {\it left} panel shows the counts map of {\it front} converted events in the energy range from 2 to 200 GeV nearby SNR {\Gg}. Units are reported in 10$^4$ counts deg$^{-2}$ with pixel binning 0.01{\degree}, and the map has been smoothed through convolution using a Gaussian distribution of width 0.2{\degree}. The {\it left} panel is a 1{\degree} $\times$ 1{\degree} image centered on {\Gg} where the magenta crosses indicate the positions of 2FGL cataloged sources in the ROI \citep{Nolan_2012} and the magenta ellipses indicate the 95$\%$ confidence limits on the position of both.  The {\it right} panel shows the corresponding TS map created for the region. The green contours show radio emission observed using the VLA at 20cm (1.4 GHz) \citep{Brogan_2006}. The black cross indicates the position of detected OH maser emission \citep{Frail_1996, Hewitt_2009} and the magenta cross within the green contours shows the position of the coincident $\gamma$-ray source 2FGL J1758.8--2402c and the VHE source HESS J1800-240C \citep{Aharonian_2008_W28, Nolan_2012}. The white curves in both panels show test statistic contours made using the TS map presented in the {\it right} panel. They correspond to detection significance measurements of 8, 9, and 10$\sigma$.}
\label{fig:Maps}
\end{figure*}

\section{OBSERVATIONS AND DATA ANALYSIS}
\label{sec:Observations}
The {\fermi} is an $e^{-} e^+$ pair conversion telescope with a field-of-view (FoV) of $\sim2.3$ sr, and an energy dependent point-spread function (PSF) that improves at higher energy. The {\fermi} is capable of detecting photons with energies from 20 MeV to 300 GeV. Further details regarding the instrument and its calibration can be found in \citet{Atwood_2009}. This work analyzes {\fermi} data spanning nearly 63 months of measurement, from 4 August, 2008 until 27 September, 2013. Only events with the Pass 7 V6 $\it{Source}$ class designation were used during this analysis, which significantly reduces the residual background rate \citep[and references therein]{Ackermann_2012}. In addition to the event class designation, analyzed events had reconstructed zenith angle values smaller than 100\degree\ to reduce the inclusion of terrestrial $\gamma$-rays \citep{Abdo_albedo}. All analyzed events came from a circular region, centered at the radio center of {\Gg}, with a radius of 25\degree. The instrument response functions (IRFs) used in the analysis correspond to the "Pass7 version 6" versions developed using in-flight data \citep{Rando_2009, Ackermann_2012}. 

Data analysis was performed using the Fermi Science Tools v9r31p1 \footnote{The Science Tools package v9r31p1 and support documents are distributed by the Fermi Science Support Center and can be accessed at http://fermi.gsfc.nasa.gov/ssc}. Maximum likelihood analysis was conducted on the data set using the $\it{gtlike}$ tool, \citep{Mattox_1996}, to study the morphological and spectral properties of the SNR. The emission models used by the fitting routines include a Galactic diffuse component describing the interactions of cosmic rays with the ISM and cosmic-ray electron interaction with interstellar radiation, and an isotropic component describing the instrumental and extragalactic diffuse backgrounds. The mapcube modeling the Milky Way $\gamma$-ray emission was the file {\tt gal\_2yearp7v6\_trim\_v0.fits}, with the isotropic emission modeled by the {\tt iso\_p7v6source.txt} table.

\subsection{Spatial Analysis}
\label{sec:Spatial}
To study the morphological characteristics of the $\gamma$-ray emission in the field of {\Gg}, events in the energy range from 2 to 200 GeV were used. Also, only {\it front}-selected photons, those converted in the first 12 scintillator planes of the {\it Fermi}-LAT Tracker, were included in the analysis to ensure better quality photon data and improved angular resolution. This gives the analyzed events a 68$\%$ containment radius angle for normal incidence photons of $\le 0.3$\degree. 

The {\it gtbin} tool was used to make counts maps of a 20\degree $\times$ 20\degree~region centered on the region of interest (ROI) defined during the event selection steps. Then the Galactic diffuse and isotropic extragalactic backgrounds were modeled and a $1\degree \times 1\degree$ test statistic (TS) map was constructed using the {\it gttsmap} tool to determine the significance of $\gamma$-ray emission detection and precisely establish the position, and possible extent, of a corresponding source. The {\it gttsmap} tool measures the test statistic across a defined grid of pixel positions. The values measured by the tool correspond to the logarithmic ratio of the likelihood of a point source existing in addition to the background model at a given point in the grid and the likelihood of the model existing without an additional source at that grid position, 2log($L_{s+b}/L_{b})$.

The ROI for SNR {\Gg} was centered at ($\alpha_{J2000} = 269.73, \delta_{J2000} = -24.06$) for the analysis. The {\it left} panel of Figure~\ref{fig:Maps} shows the 1{\degree} $\times$ 1{\degree} countsmap made using the {\it gtbin} tool for the 2 -- 200 GeV energy band, with a pixel binning of {0.01\degree}. This map have also been smoothed using a Gaussian distribution of width {0.2\degree}. Overlaid on the {\it left} panel are test statistic (TS) contours made using the TS map created with the {\it gttsmap} tool and shown in the {\it right} panel of the figure. To create the TS map, four sources of $\gamma$-ray emission were incorporated into the background source model. These included the two sources of diffuse $\gamma$-ray background, both the Galactic and extragalactic components, the $\gamma$-ray emission from a nearby $\gamma$-ray pulsar, PSR J1800--240, which is coincident with both 2FGL J1800.8--2400 and the VHE $\gamma$-ray source HESS J1800--240B, and lastly the extended $\gamma$-ray emission associated with W28, modeled in the 2FGL catalog by source J1801.3--2326e, which dominates the emission in the northeast corner of the countsmap. 2FGL J1800.8--2400 was modeled as a point source with a power law spectral distribution while 2FGL J1801.3--2326e was modeled as a disk with a log parabola distribution and a $0.39{\degree}$ radial extension, first described by \citep{Abdo_2010_W28}. The parameters for both sources were fixed during the analysis at the values reported in the 24-month {\it Fermi}-LAT Second Source Catalog \citep{Nolan_2012} \footnote{The data for the 1873 sources in the {\it Fermi} LAT Second Source Catalog is made available by the Fermi Science Support Center at http://fermi.gsfc.nasa.gov/ssc/data/access/lat/2yr\_catalog/}. The vertical scale shows the measured TS values at each pixel position in the map, where $\sqrt{\rm TS}$ is approximately the significance in units of the standard deviation. Using these values, contours were made that represent significant emission at 8, 9, and 10$\sigma$, given in white. 

Both panels also show a black cross representing the location of OH(1720 MHz) maser detection and overlaid radio contours, in green, from a 1.4 GHz VLA observation of {\Gg} \citep{Brogan_2006}. 
The location of {\FGL} is coincident with the position of VHE  $\gamma$-ray source, HESS J1800--240C, and is marked by the magenta cross interior to the green radio contours \citep{Aharonian_2008_W28}. Finally, the magenta ellipses in both panels show the 95$\%$ confidence limit on the position of the two 2FGL $\gamma$-ray point sources mentioned above \citep{Nolan_2012}.

\subsection{Spectral Analysis}
\begin{figure}[h]
\centering 
\includegraphics[width=1\linewidth]{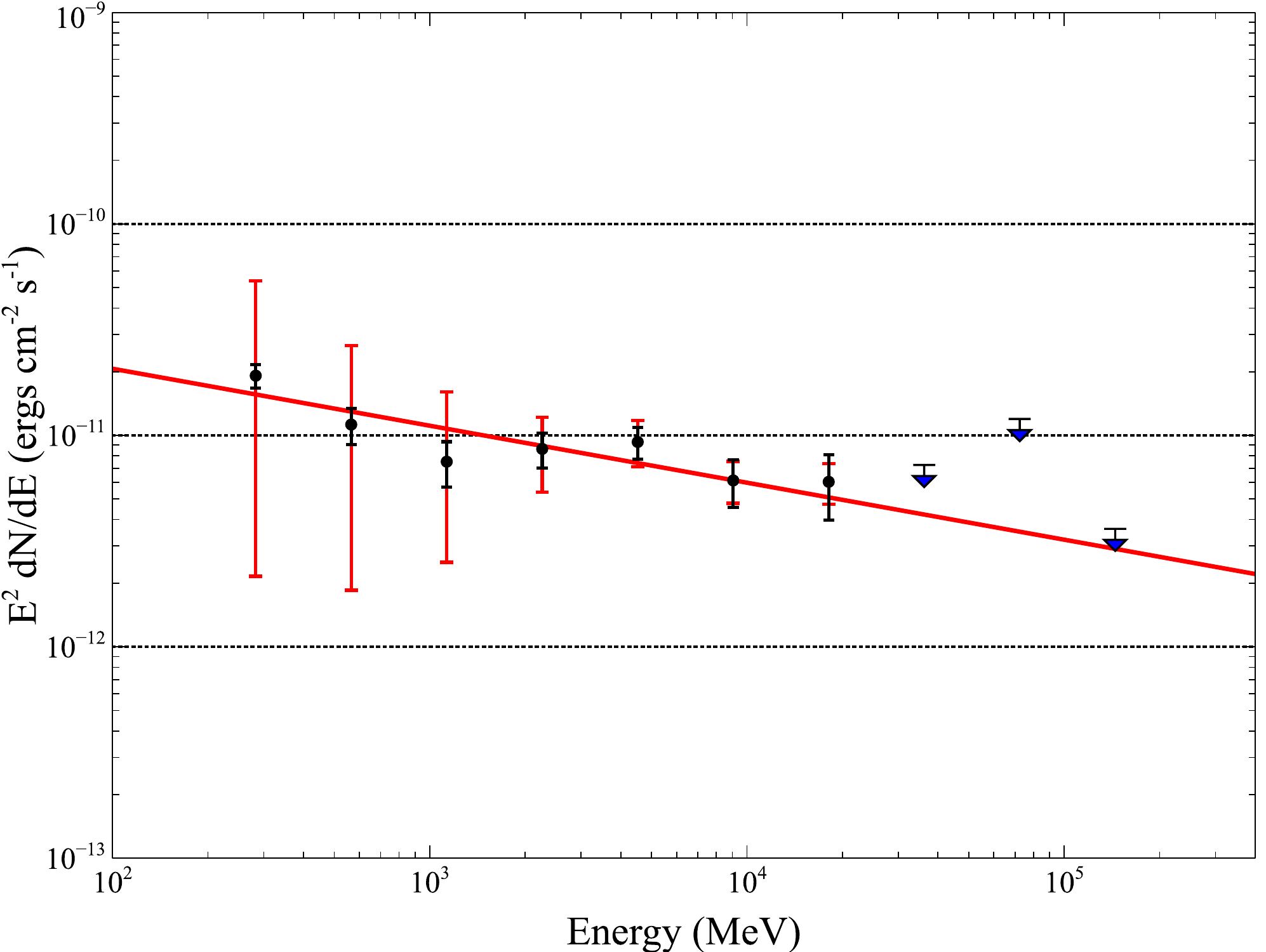} 
\caption{Logarithmic plot of the high energy $\gamma$-ray SED measured for SNR {\Gg}. The data points (black circles) represent flux values produced by the likelihood maximization process for energy ranges with $>$5$\sigma$ detections, TS $>25$. The blue errorbars indicate statistical uncertainties resulting from the likelihood fitting while the red errorbars are systematic errors. The inverted blue triangles correspond to calculated upper limits set on the energy flux for energy ranges with TS $<25$. Included is a power law fit to the significant data (red line) giving a measure of the spectral index, $\Gamma=2.27 \pm 0.12$.} 
\label{fig:Results} 
\end{figure}

To study the $\gamma$-ray spectral energy distribution (SED) of {\Gg}, likelihood analysis was performed using both {\it front} and {\it back} converted events for the energy range 0.2--204.8 GeV. The lower energy value is selected to avoid the highly variable effective area of the instrument at low energies and because of the large uncertainty of the Galactic diffuse model for energies below 0.2 GeV. The {\it gtlike} tool is used to model the $\gamma$-ray flux for logarithmically spaced energy bins and to estimate the best-fit spectral parameters through likelihood maximization fitting. This requires modeling both the Galactic diffuse and isotropic extragalactic background and all $\gamma$-ray point sources within 25{\degree} of the source of interest. The point sources are modeled in accordance with the 2FGL catalogue \citep{Nolan_2012}. The parameters of the galactic and extragalactic background models and all 2FGL sources within 5{\degree} of the source of interest were allowed to vary during the fitting. This included the parameters of the extended $\gamma$-ray source 2FGL J1801.3--2326e which was modeled in the same way as in Section~\ref{sec:Spatial}. 
The modeled flux values resulting from the likelihood analysis are reported in Figure~\ref{fig:Results}. The statistically significant flux points represent the energy range 0.2--25.6 GeV. The black error bars represent statistical uncertainties associated with the likelihood fitting method. The red error bars represent the systematic uncertainties. For the IRFs used in this analysis, the systematic uncertainties associated with the effective area is energy dependent. For 100 MeV the uncertainty is 10$\%$, decreasing to 5$\%$ at 560 MeV, and increasing to 20$\%$ at 10 GeV \citep[and references therein]{Abdo_103_2009, Nolan_2012}. Additionally, the uncertainty in the flux of the Galactic diffuse background is also considered. This was done by artificially varying the normalization of the Galactic background model from the best-fit flux values by $\pm6\%$ for each energy bin. Using the artificially adjusted background values and freezing them, we remeasured the flux values and used the new measurements to approximate the upper and lower limits of the uncertainty associated with the Galactic diffuse background. This treatment is similarly applied in \citet{Abdo_706_2009, Castro_2010}; and \citet{Castro_2013}.

For the energy ranges with measured TS values  $<25$, the recorded $\gamma$-ray flux has a significance of less than 5$\sigma$. In these cases, the flux values are not reported but instead replaced by an upper limit measured using the python based likelihood analysis tool, {\it pyLikelihood}. These flux upper limits are presented in Figure~\ref{fig:Results} alongside the flux measurements as inverted triangles. Also included is a $\chi^2$-fit power law distribution for the seven significant flux values. The fitting measured a spectral index of $\Gamma = 2.27 \pm 0.12$. Integrating this flux model over the energy range, an estimate of the $\gamma$-ray flux for the energy range between 100 MeV and 100 GeV measured at $F_{\gamma} \approx 6.46 \times 10^{-11}$ erg/cm$^2$/s. Using both kinematic distance measurements indicated by the adjacent OH(1720 MHz) maser emission velocity, 3.1 and 13.7 kpc, the luminosity over this band for each distance is estimated at $L_{\gamma, 3.1} \approx 7.4 \times 10^{34}$ erg/s and $L_{\gamma, 13.7} \approx 1.45 \times 10^{36}$ erg/s.

\subsection{XMM-Newton Analysis}
\label{sec:XMM}
G5.7$-$0.1 was observed by XMM-Newton for $\sim 32$~ks on 16 September 2012 (ObsID 0691990101). The MOS and pn cameras were operated in full frame mode, and the "medium" optical filter was used. The data were reduced with the XMM-SAS software package (version 12.0.1) with calibration files current through XMM-CCF-REL-296, using the standard procedure for reducing XMM-Newton data outlined in the XMM-Newton ABC Guide and the Birmingham XMM-Newton Guide.

\begin{figure*}[ht]
\centering
\includegraphics[width=0.95\linewidth]{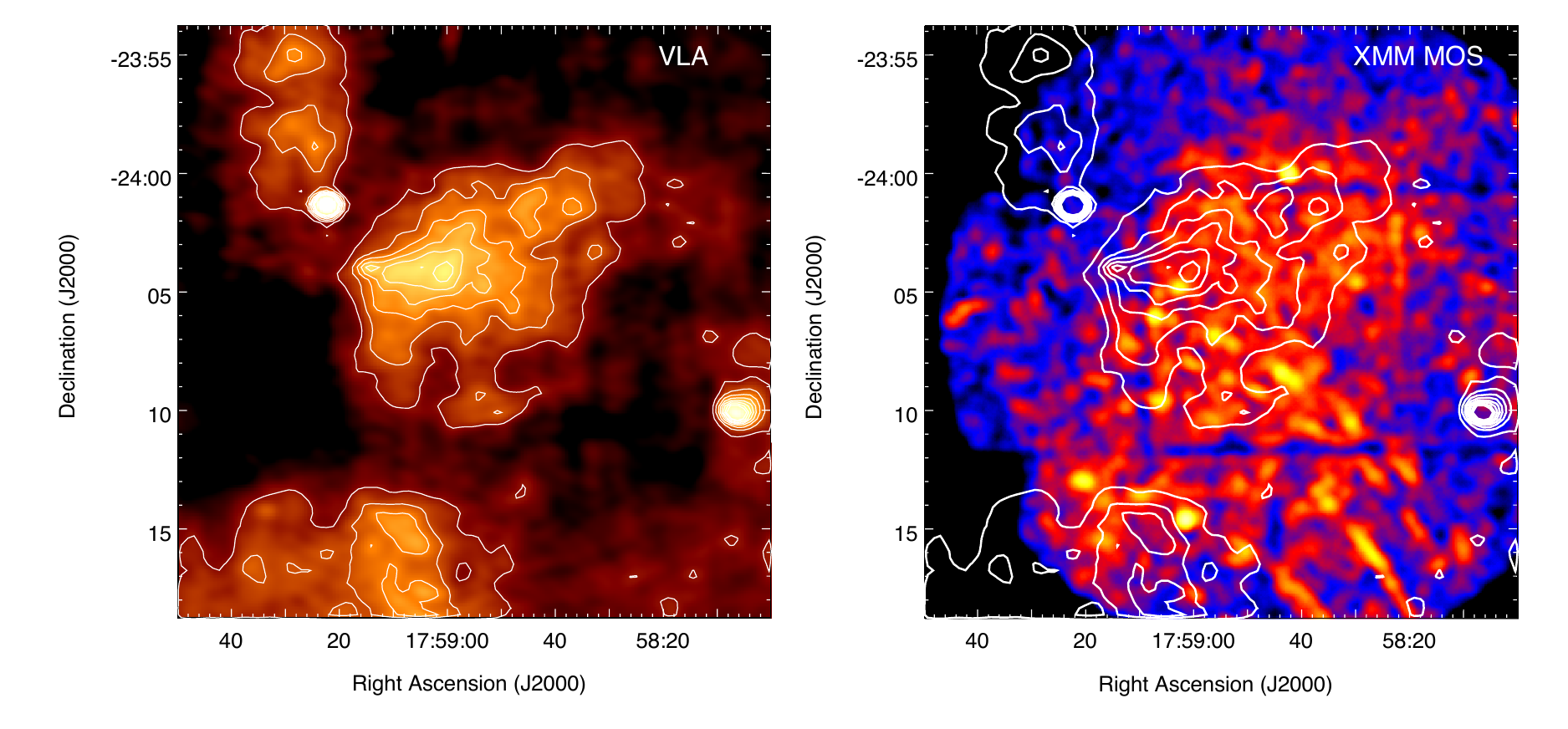}
\caption{Radio (left) and X-ray (right) images of G5.7$-$0.1. Contours correspond to the 1.4~GHz VLA intensity (using a linear scale from 1 to 20 mJy~${\rm\ beam}^{-1}$).  The arc-like stripes in the lower half of the X-ray image are stray-light artifacts produced by singly-reflected X-rays from the bright off-axis source GX~5$-$1, located to the southeast of the SNR (and outside of the telescope field of view).}
\label{fig:XMM}
\end{figure*}

The MOS2 image of G5.7$-$0.1 is shown in Figure~\ref{fig:XMM} ({\it right}). The image has been filtered to the energy range 0.5 - 2.6 keV and smoothed by a 36 arcsec Gaussian.  Distinct arc-like artifacts can be seen in the lower half of the image. These result from stray light associated with singly-reflected X-rays from the bright off-axis low mass X-ray binary GX~5$-$1.  The artifacts are exceptionally bright at higher energies, but modest below 2.6~keV due to the high column density for GX~5$-$1. Contours are from the 1.4~GHz VLA image shown in the {\it left} panel of Figure~\ref{fig:XMM} \citep{Brogan_2006}.

We extracted spectra from the northern ``stripe-free'' half of G5.7$-$0.1 and found that the emission is adequately described by an absorbed solar-metallicity plasma with $kT \sim 0.65$~keV and $N_H = (1.3 \pm 0.1) \times 10^{22}{\rm\ cm^{-2}}$, presumably associated with the SNR, accompanied by an absorbed bremsstrahlung component with $kT = 7.2$~keV and $N_H = 2.5 \times 10^{22}{\rm\ cm^{-2}}$, consistent with scattered emission from GX~5$-$1 \citep{Smith_2006}. The contribution from the latter component prevents detection of any potential nonthermal emission from G5.7$-$0.1.

As noted in Section~\ref{sec:Introduction}, OH(1720 MHz) maser measurements place G5.7$-$0.1 at a distance of either $\sim 3.1$~kpc or $\sim 13.7$~kpc. The estimated distance to GX~5$-$1 is $\sim 8$kpc \citep{Smith_2006}. The lower column density for G5.7$-$0.1 indicates that it is closer than GX~5$-$1, suggesting that the distance ambiguity is broken in favor of the near distance.

\section{Discussion}
\label{sec:Discussion}
Based in part on the observation of nonthermal X-ray and $\gamma$-ray emission, SNRs are known to accelerate particles to very high energies. These accelerated particles produce high energy emission through either leptonic processes, including nonthermal bremsstrahlung and inverse Compton scattering of ambient photons by accelerated electrons, or through the decay of neutral pions resultant from the collision of accelerated hadrons with ambient protons. To better understand the nature and origin of the broadband, nonthermal emission in the direction of {\Gg}, it was necessary to model it using methods described in depth by \citet{Castro_2013}. 

\subsection{Modeling broadband emission}
Models simulating emission due to shock acceleration of electrons and protons were used to match the broadband SED of {\Gg} particularly the observed $\gamma$-ray emission. To analyze the $\gamma$-ray SED of the accelerated particles, and assuming isotropy of both particle species, proton or electron, the spectral momentum distribution of the accelerated particles, $dN_i/dp$, was modeled using a power law with an exponential cutoff,
\begin{equation}
\frac{dN_i}{dp} = a_i p^{-\alpha_i} \exp\left(-\frac{p}{p_{0,i}}\right)
\label{eq:distribution}
\end{equation}
where $i$ denotes the particle species, and $p_{0,i}$ is the corresponding exponential cutoff momentum for that species \citep{Castro_2010}. Transforming the particle distribution to energy space allows the exponential cutoff to be defined by an input energy, $E_{0,i}$. The parameter $\alpha$ is the particle momentum distribution index. The spectral index is related to $\alpha$ as $\Gamma = \alpha - 2$. For the case of re-acceleration of ambient cosmic rays, the index values for the electrons and protons differ at the outset, but reach the same values as the process approaches the steady state \citep{Tang_2015}. Here we assume $\alpha_p = \alpha_e.$ The ratio of the normalization constants of these particle distributions, $a_i$, establishes the electron to proton ratio, $k_{ep} \equiv a_e/a_p$, a value of particular interest to this analysis. These coefficients are implicitly determined by defining $k_{ep}$ during the modeling and setting the total integrated energy converted into particle acceleration within the SNR shell $E_{CR} = \theta E_{SN}$, where $\theta$ represents the efficiency of converting the total SN energy into shock acceleration of cosmic rays. The total mechanical energy for the SN was set to the canonical value, $E_{SN} = 10^{51}$ erg. The model assumes that the SNR is a spherical shell of shock-compressed material, with thickness $R/12$ and compression ratio 4, expanding into a uniform medium of density $n_0$ \citep{Castro_2013}. This ratio is expected to be somewhat larger if efficient particle accelerating is in fact occurring \citep{Warren_2005, Cassam_2008}.

Modeling of the $\pi^{0}$-decay emission mechanism is based on predictions from \citet{Kamae_2006}, while the synchrotron and inverse Compton scattering (IC) emission components utilize models described by \citet{Baring_1999}. The inverse Compton scattering model assumes an ambient photon field of Cosmic Microwave Background (CMB) photons with $kT_{CMB} = 2.725$ K. Finally, the nonthermal bremsstrahlung emission is modeled using the work of \citet{Bykov_2000}. Further description of the model used in this work can be found in \citet{Castro_2013} and references therein. 

Several parameters were adjusted to model the observed broadband spectrum for both the leptonic and hadronic scenarios. These included the momentum distribution index, $\alpha$, with 4.00 representing a flat spectral shape, the postshock magnetic field strength, $B_2$, the ambient density, $n_0$, and the cutoff energies, $E_{0,i}$. An adjustment of $B_2$ affects the flux of the radio synchrotron model. The nonthermal bremsstrahlung and $\pi^0$-decay components of the high energy model are directly influenced by changing the value of $n_0$. During the modeling, the cutoff energies were constrained to $E_0,e \leq E_0,p$ to account for the differing radiative loss scenarios of accelerated electrons and protons.

\begin{figure}[h]
\centering
\includegraphics[width=1\linewidth]{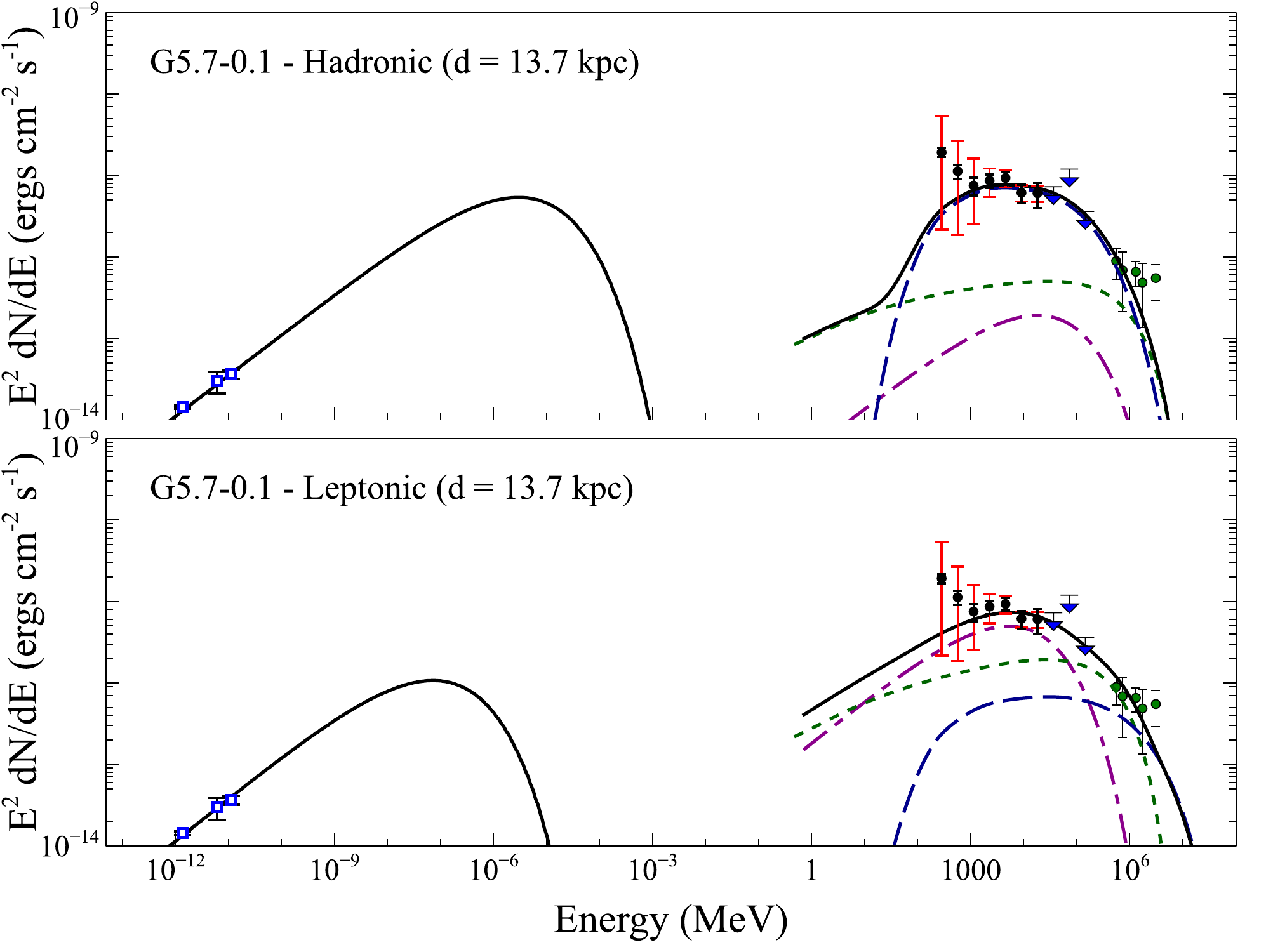}
\includegraphics[width=1\linewidth]{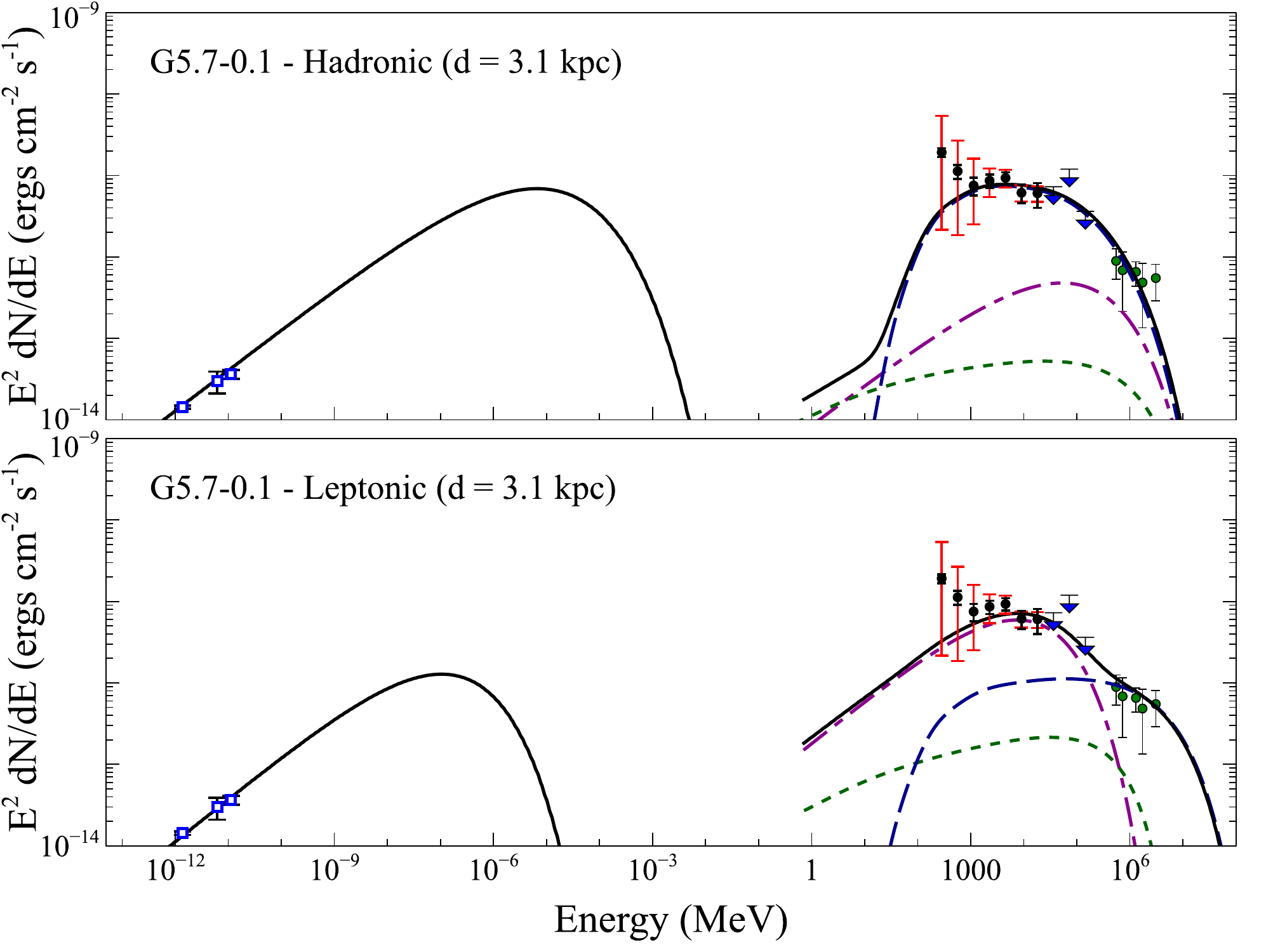}
\caption{Broadband fits to radio observations (blue circles) \citep{Brogan_2006}, and $\emph{Fermi}$-LAT observations (black circles) of SNR {\Gg} with SNR distance assumed at 13.7 kpc (top) and 3.1 kpc (bottom). Flux points for the VHE $\gamma$-ray source HESS J1800--240C are included as green circles \citep{Aharonian_2008_W28}. The solid black curve represents the total modeled nonthermal emission. Also shown are the constituent spectra from IC (magenta; dot-dashed), $\pi^0$-decay (blue; dashed), and nonthermal bremsstrahlung (green; dotted) emission. The lower energy component modeled the radio synchrotron emission.}
\label{fig:both}
\end{figure}

\begin{table*}
\centering
\caption{Inputted Model Parameters}
\begin{tabular}{cccccccccc} 
\hline\hline 
d & Model & $\alpha$ & E$_{0,p}$  & E$_{0,e}$ & $k_{ep}$ & $n_0$  & $B_2$  & $E_{CR,p}$ & $E_{CR,e}$  \\ [0.5ex] 
(kpc) & & &(TeV) & (TeV) & & (cm$^{-3}$) & ($\mu$G) & (10$^{50}$erg) & (10$^{50}$erg)   \\
\midrule
 13.7&$\emph{Hadronic}$ & 4.00 & 2.0 & 2.0 & 0.01 & 22.0 & 17.0 & 3.894 & 0.116   \\
 &$\emph{Leptonic}$ & 3.95 & 15.0 & 1.0 & 0.5 & 2.0 & 1.5 & 4.702 & 2.298   \\
\midrule 
3.1&$\emph{Hadronic}$ & 4.00 & 2.4 & 2.4 & 0.001 & 1.2 & 12.0 & 3.993 & 0.007   \\
&$\emph{Leptonic}$ & 3.95 & 40.0 & 1.2 & 0.035 & 0.22 & 1.5 & 3.881 & 0.118   \\\hline
\hline
\end{tabular}
\label{tab:Model}
\end{table*}

\begin{figure*}
\centering

\includegraphics[width=0.84\linewidth]{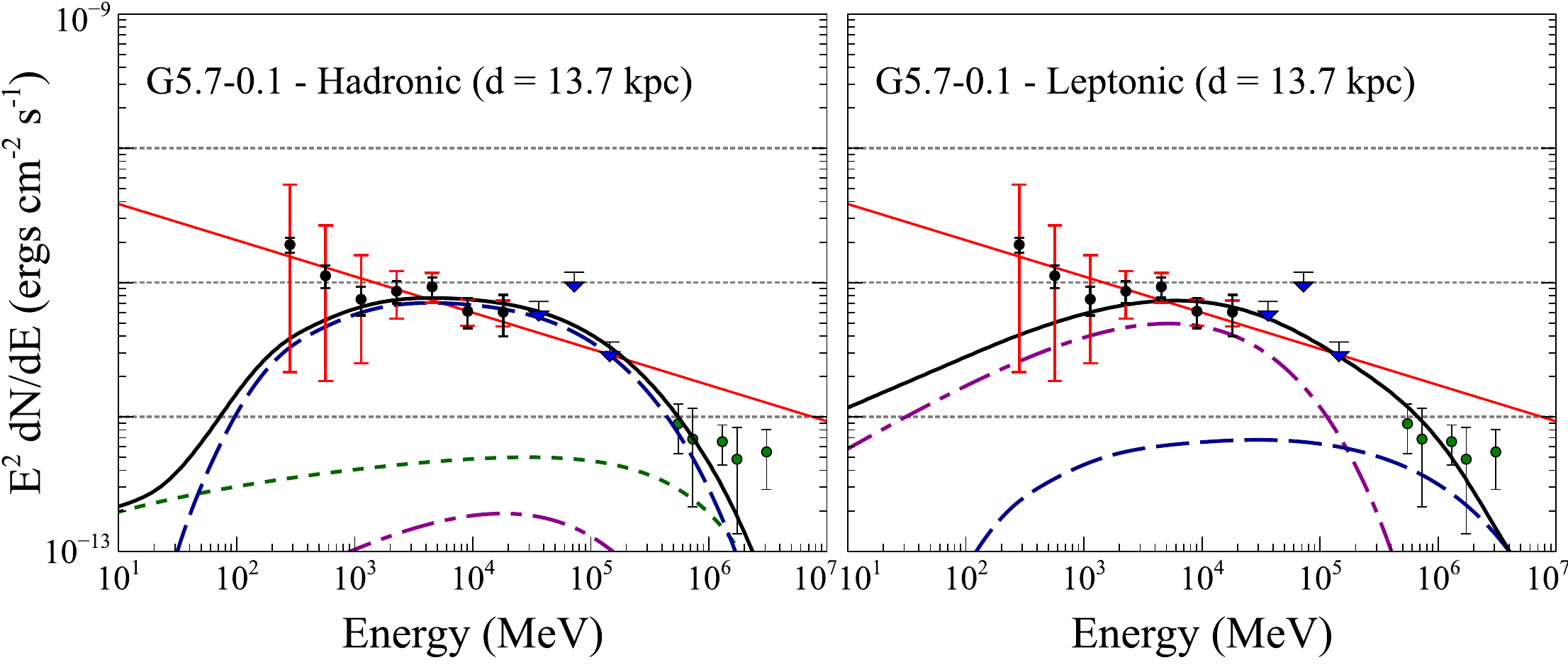}

\includegraphics[width=0.84\linewidth]{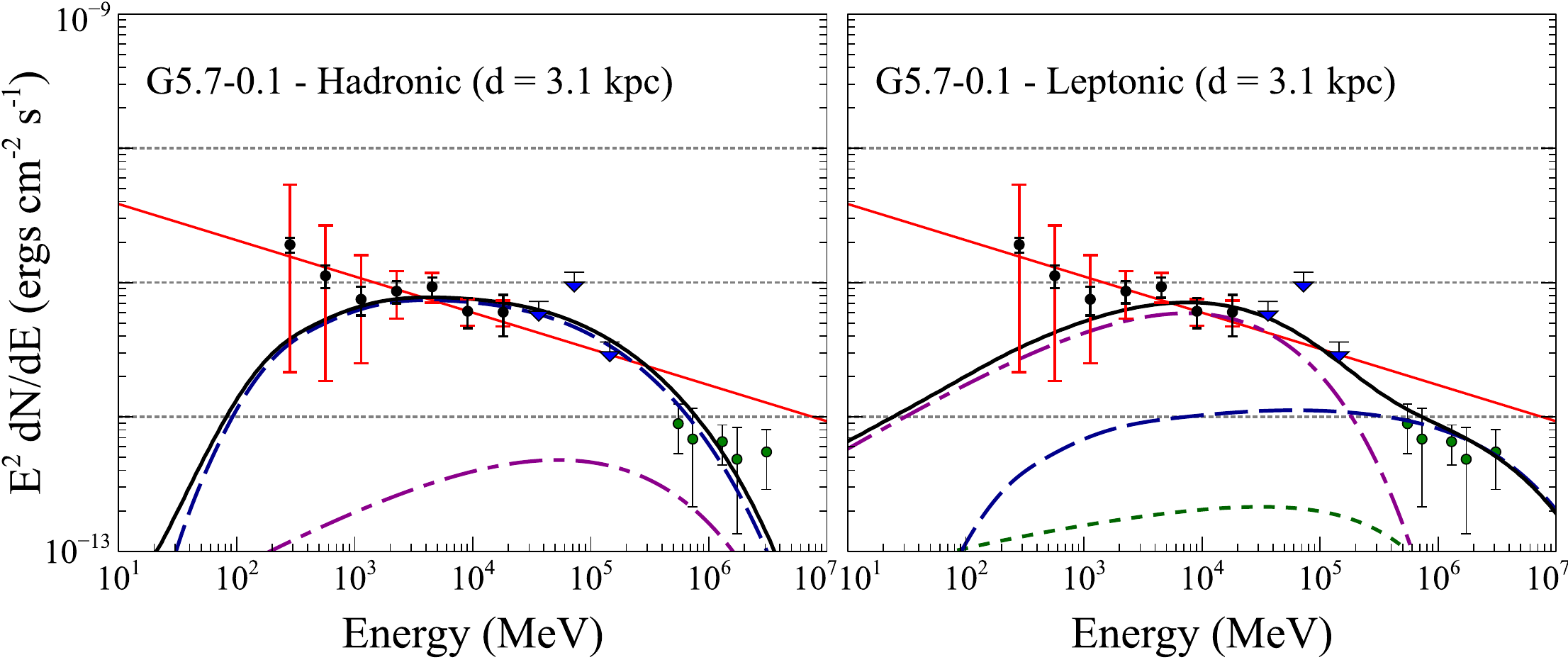}

\caption{Results of both hadronic (left) and leptonic (right) model fitting of $\emph{Fermi}$-LAT detected $\gamma$-ray emission (black circles) and calculated upper limits (blue triangles) with SNR distance at 13.7 kpc (top) and 3.1 kpc (bottom). Flux points for the VHE $\gamma$-ray source HESS J1800--240C are also included as green circles \citep{Aharonian_2008_W28}. The red errorbars are systematic uncertainties and the black errorbars are the statistical uncertainties. The red line represents a power law model fit to the data points covering the range 0.2-25.6 GeV. The solid black represents the total nonthermal emission as predicted by the model. The constituent modeled spectra from inverse Compton emission (magenta; dot-dashed), $\pi^0$-decay (blue; dashed), and nonthermal bremsstrahlung (green; dotted), are also shown.}
\label{fig:bothzoom}
\end{figure*}

The electron to proton ratio has been measured by observations of cosmic rays (CRs) at Earth's surface to be $k_{ep} = 0.01$ \citep{Hillas_2005}. There is also a known difficulty in injecting electrons into the DSA process. This means it is quite reasonable to expect $k_{ep} < 1$ and potentially $k_{ep} <<1$ at the particle acceleration site \citep{Park_2015}. \citet{Castro_2013} demonstrates that at $k_{ep} = 0.01$ the hadronic component will dominate over the IC emission component for ambient densities larger than $\sim1$ cm$^{-3}$ over the energy range used to observe {\Gg} in this study. It was shown that the model predicts hadronic emission to dominate over the nonthermal bremsstrahlung for all of the particle distributions considered by \citet{Castro_2013} as well. Therefore, the ratio was loosely constrained around $k_{ep} = 0.01$ during the $\pi^0$-decay dominated modeling. It was also shown that an electron-proton ratio larger than $k_{ep}\sim 0.1$ is required for the nonthermal bremsstrahlung component to dominate the $\gamma$-ray emission in the {\fermi} band. Because of these considerations, nonthermal bremsstrahlung is not expected to dominate the applied models \citep{Castro_2013}. 

The final constraint fixed the total energy of the accelerated particles in the hadronic models to $E_{CR} \equiv E_{CR,p} + E_{CR,e} = 4 \times 10^{50}$ erg. This assumes that 40$\%$ of the SNR energy from a 10$^{51}$ erg supernova is deposited into relativistic particles. This represents a reasonable upper limit on this energy and therefore sets a lower limit on the ambient density, $n_0$, which, as mentioned previously, is constrained during the modeling. This should allow comparison between the estimated density predicted by the modeling with expectations for shocked ISM. The total CR energy and $k_{ep}$ are varied in the inverse Compton dominated scenario to fit the measured flux while limiting the contribution of the hadronic component over the {\fermi} energy range \citep{Castro_2013}. 

\subsection{SNR {\Gg} emission}
Due to the ambiguity in the measured distance to {\Gg}, broadband modeling of the $\gamma$-ray SED and radio observations of the SNR was conducted for both potential distances reported by \citet{Hewitt_2009}. Figure~\ref{fig:both} shows the broadband fits for both hadronically and leptonically dominated emission scenarios for the SNR being positioned at 13.7 kpc (top) and 3.1 kpc (bottom). The radio emission was measured by \citet{Brogan_2006} using VLA observations of the SNR at 90 cm (333 MHz), 20 cm (1.49 GHz) and 11 cm (272 MHz). These values and the {\fermi} flux points were fit by both the $\pi^{0}$-decay and IC-dominated emission models described in the previous section. The flux of VHE $\gamma$-ray source HESS J1800--240C, believed to be associated with {\Gg}, is included as green circles \citep{Aharonian_2008_W28}. The results of the model fitting are shown in Table~\ref{tab:Model}. Figure~\ref{fig:bothzoom} focuses on the highest energy range of the spectrum in order to better understand the differences between the resulting fit of the two models at each distance. Also included in Figure~\ref{fig:bothzoom} is the simple power law model fit applied previously and shown in Figure~\ref{fig:Results}. 

Applying a model dominated by the hadronic emission mechanism to the $\gamma$-ray emission in the direction of SNR {\Gg} was favored at both distances. At a distance of 13.7 kpc, the hadronic model used the electron-proton ratio observed on Earth, $k_{ep}=0.01$ with a spectral index of $\Gamma=2.0$, and a total energy of accelerated particles set to $4 \times 10^{50}$ erg. Additional parameters were adjusted until the model closely approximated the radio and $\gamma$-ray flux measurements, including the HESS observations, and respected the calculated upper limits. Modeling indicated cutoff energies equal to $E_{0,i}=2.0$ TeV and an ambient density estimate of $n_{0,13.7}=22.0$ cm$^{-3}$. Also a magnetic field of 17.0 $\mu$G was required to accurately model the radio synchrotron flux. The resulting estimation of the energy converted into the accelerating cosmic ray protons (CRPs) was $\sim 3.89 \times 10^{50}$ erg, with the remaining $\sim 0.11 \times 10^{50}$ erg used in cosmic ray electron (CRE) acceleration. These results are consistent with previous studies of SNR-MC systems that confirmed a hadronically dominated emission mechanism \citep{Castro_2013, Auchettl_2014}

The leptonic model at a distance of 13.7 kpc also closely approximated the measured data points and upper limits values. However, the model required an electron-proton ratio of $k_{ep} = 0.5$, significantly larger than that suggested by ground observation. Using this ratio and a spectral index of $\Gamma = 1.95$, different cutoff energies for both particle species were required to model the SED. Below $\sim 20$ GeV the leptonic components dominated the spectrum while above this energy range, the pion-decay channel was extended in order to better model the HESS VHE detection. The cutoff energies for the electron and proton components were $E_{0,e}=1.0$ TeV and $E_{0,p}=15.0$ TeV respectively. An ambient density of $n_{0,13.7}=2.0$ cm$^{-3}$, and a magnetic field of 1.5 $\mu$G were also required to adequately approximate the SED. The data required an increase in the total supernova energy from the canonical value resulting in an estimated energy used to accelerate CRPs of $\sim 4.7\times 10^{50}$ erg and $\sim 1.5\times10^{50}$ erg to accelerate CREs. These are somewhat higher values that would require an SN having both efficient particle acceleration and slightly more energy than is typically expected. Several factors including a relatively high proton cutoff energy, a slightly higher required $E_{SN}$, and a notably high required electron-proton ratio, provide little support favoring a leptonically dominated emission scenario.

Applying the hadronically dominated emission model to the broadband SED while setting an SNR distance of 3.1 kpc required an electron-proton ratio of $k_{ep} = 0.001$, an order of magnitude lower than that observed on Earth, and a hard spectral index of $\Gamma=2.0$. A lower $k_{ep}$ such as this has also been seen in the modeling of the $\gamma$-ray SEDs of SNRs like RX J1713.7--3746 and W41 \citep{Ellison_2010, Castro_2013}. This value resulted in a total energy deposited in CRP acceleration of $\sim 3.97 \times 10^{50}$ erg and $\sim 0.03 \times 10^{50}$ erg used to accelerate CREs. A magnetic field of $B_2 = 4.0 ~\mu$G was required to approximate the radio synchrotron data. An important difference between the model parameters for the hadronic scenarios at the two distances were the required ambient densities. For an SNR distance of 3.1 kpc the model required a value of $n_{0,3.1} = 1.0$ cm$^{-3}$, an order of magnitude smaller than that required for the 13.7 kpc model. These parameters are all quite reasonable and provide convincing evidence of a hadronic origin to the observed $\gamma$-ray emission.

The leptonically dominated model also fit the broadband SED adequately. The model used a spectral index of $\Gamma=1.95$ and required an electron to proton ratio of $k_{ep} = 0.035$. This somewhat higher $k_{ep}$ was required to keep the IC contribution the dominant component of the total $\gamma$-ray flux below $\sim 20$ GeV. Again two separate cutoff energies were required to accurately model the HESS VHE detection. These cutoff energies were $E_{0,e}=1.2$ TeV and $E_{0,p}=40.0$ TeV respectively. Setting an ambient density of $n_{0,3.1} = 0.22$ and a magnetic field of 1.5 $\mu$G resulted in a strong fit to the SED. These parameters resulted in an estimated CRP acceleration energy of $\sim 3.88 \times 10^{50}$ erg and $\sim 0.12 \times 10^{50}$ erg converted to CRE acceleration. It is important to point out that while the leptonic component dominates the model, the required pion-decay component is quite significant. Therefore, it appears that at this SNR distance a significant leptonic component to the $\gamma$-ray emission cannot completely be ruled out. However in both the hadronic and leptonically dominated modeling a significant hadronic component was required.

Further analysis was carried out to see if additional photon fields could bolster the IC component enough that a strong fit to the data was possible while using an electron-proton ratio of $k_{ep} = 0.01$ and a total energy in accelerated particles of $E_{CR} = 4 \times 10^{50}$ erg. A photon field of temperature $T=25$K, which is characteristic of interstellar dust, was used in this modeling. For an SNR distance of 13.7 kpc, an additional photon field with a density $\sim 40$ times that of the CMB was required. This is a significantly higher density than what is expected from a similar Galactic photon field \citep{Strong_2000}. At 3.1 kpc an additional photon field $\sim 5.1$ times the density of the CMB was necessary to fit the SED. This resulted in a strong fit with $n_{0,3.1}=0.43$ cm$^{-3}$, $B_2 = 2.5$ $\mu$G and reasonable acceleration energies, $E_{CR,p} = 3.94 \times 10^{50}$ erg and $E_{CR,e} = 0.06 \times 10^{50}$ erg. This photon density is again too large for the region, leaving the hadronic dominated model the most convincing fit to the data at either distance. 

While a hadronically dominated model appear to be the most appropriate fit to the broadband SED at either distance, the lower column density measured during the XMM-Newton analysis of {\Gg}, described in Section~\ref{sec:XMM}, makes a strong argument in favor of the 3.1 kpc distance. The parameter values required to model the SED at 3.1 kpc match closely to previous studies of similar SNR-MC systems that used this model \citep{Castro_2010, Slane_2012, Castro_2012, Castro_2013, Ackermann_2013}. Therefore it seems reasonable to conclude that {\Gg} is an independent and significant SNR-MC $\gamma$-ray source, located at 3.1 kpc. 

\section{Summary}
The work presented herein studied the spatial and spectral characteristics of the $\gamma$-ray emission from the region surrounding the SNR {\Gg} that is known to be interacting with nearby molecular clouds. This study found a significant region of $\gamma$-ray emission in the direction of the SNR at 10$\sigma$ for the energy range 2--200 GeV with the $\it{Fermi}$-LAT. We also successfully measured the $\gamma$-ray emission spectrum for the 0.2--204.8 GeV band, detecting significant flux values for the energy range 0.2--25.6 GeV. The radio and $\gamma$-ray broadband emission spectrum was then fit to a model for emission from a cut-off power law distribution of relativistic protons and electrons. This modeling allowed conclusions to be drawn about the origin of the observed emission. In an effort to set constraints on the position of the SNR, the model was applied while assuming two different potential SNR distances. In both cases the broadband SED was better explained by a model that assumed a hadronic origin of the $\gamma$-ray flux than one assuming a leptonic dominated scenario. XMM-Newton observations also measured a column density for the SNR that supported an SNR distance of $\sim 3$ kpc. This potentially establishes SNR {\Gg} as an SNR-MC system independent of the SNR W28 cloud complexes and a significant $\gamma$-ray source.
\label{sec:Summary}

\acknowledgments
DC acknowledges support for this work provided by the National Aeronautics and Space Administration through the Smithsonian Astrophysical Observatory contract SV3-73016 to MIT for Support of the Chandra X-Ray Center, which is operated by the Smithsonian Astrophysical Observatory for and on behalf of the National Aeronautics Space Administration under contract NAS8-03060. PS acknowledges support from NASA Contract NAS8-03060. JDG acknowledges support for analysis of the XMM-Newton data through the National Aeronautics Space Administration under Grant NNX13AD56G. The authors would also like to thank the helpful comments of the referee.


\bibliography{bibliography}

\end{document}